\documentclass[preprint,review,12pt]{elsarticle}

\usepackage{xcolor}
\usepackage{color}
\usepackage{soul}

\usepackage{lineno,hyperref}
\usepackage{graphicx}
\usepackage{amssymb}
\usepackage{lineno}
\usepackage{float}
\usepackage{multirow}
\modulolinenumbers[5]

\journal{Carbon}

\begin{document}

\begin{frontmatter}

\title{Tensile properties of carbon nanotube fibres described by the fibrillar crystallite model}

\author[Technical University of Madrid,IMDEA Materials Institute]{Juan C.  Fern\'{a}ndez-Toribio}

\author[IMDEA Materials Institute]{Bel\'{e}n Alem\'{a}n}

\author[Technical University of Madrid]{\'{A}lvaro Ridruejo \corref{correspondingauthors}}

\cortext[correspondingauthors]{Corresponding authors}
\ead{alvaro.ridruejo@upm.es}

\author[IMDEA Materials Institute]{Juan J.  Vilatela \corref{correspondingauthors}}

\ead{juanjose.vilatela@imdea.org}

\address[Technical University of Madrid]{ Department of Materials Science
Polytechnic University of Madrid, 28040, Madrid (Spain)}
\address[IMDEA Materials Institute]{ IMDEA Materials Institute. Eric Kandel, 2, Tecnogetafe, 28906, Getafe, Madrid (Spain)}

\begin{abstract}
This work presents a model that successfully describes the tensile properties of macroscopic fibres of carbon nanotubes (CNTs). The core idea is to treat the fibres as a network of crystallites, similar to the structure of high-performance polymer fibres, with tensile properties defined by the crystallite orientation distribution function (ODF), shear modulus and shear strength. Synchrotron small-angle X-ray scattering measurements on individual fibres are used to determine the initial ODF and its evolution during \textit{in-situ} tensile testing. This enables prediction of tensile modulus, strength and fracture envelope, with remarkable agreement with experimental data for fibres produced in-house with different constituent CNTs and for different draw ratios, as well as with literature data. The parameters extracted from the model include: crystallite shear strength, shear modulus and fibril strength. These are in agreement with data for commercial high-performance fibres, although high compared with values for single-crystal graphite and short individual CNTs. The manuscript also discusses the unusually high fracture energy of CNT fibres and exceptionally high figure of merit for ballistic protection. The model predicts that small improvements in orientation would lead to superior ballistic peformance than any synthetic high-peformance fiber, with values of strain wave velocity ($U^{1/3}$) exceeding $1000 m/s$.
\end{abstract}

\end{frontmatter}


\section{Introduction}

Carbon nanotubes (CNT) remain one of the most interesting nanobuilding blocks currently available for macroscopic applications. They can be produced in large quantities as a highly graphitic material with well defined structure and surface chemistry, in some cases with control over their molecular composition in terms of number of layers, diameter and chiral angle. When assembled into macroscopic fibres, the natural embodiment for a one-dimensional material, they have led to materials on par or stronger than conventional high-performance fibres \cite{behabtu2008carbon, koziol2007high, liu2014polymer}, higher thermal conductivity than copper \cite{gspann2017high, behabtu2013strong}, and higher mass-normalised electrical conductivity than most metals \cite{lekawa2014electrical, behabtu2013strong}. Several other applications in energy storage \cite{senokos2017large} and optoelectronic devices exploit their large specific surface and bending compliance. These examples give testimony of the efficient exploitation of the properties of individual CNTs on a macroscopic scale. 

Nevertheless, the development of theoretical models able to successfully describe the physical properties of CNT fibres as a function of their structure has proved an elusive challenge.  Hence, bulk properties of CNT fibres are still largely optimized by trial-and-error. A fundamental difficulty arises because of the inherently complex hierarchical structure of CNT fibres  \cite{vilatela2010yarn}. Such complexity stems from the confluence of many parameters determining bulk properties, including those linked with the physical and chemical properties of constituents (number of layers in CNTs, chiral angle, diameter, presence of impurities, etc), their spatial arrangement (orientation, bundle formation) and interaction between building blocks. 

In the context of mechanical properties, comparison of different CNT fibres has led to some agreement on the qualitative effects of different structural features. Higher CNT alignment parallel to the fibre axis was early identified as key to obtain high tensile strength and stiffness \cite{koziol2007high, aleman2015strong, QingwenBradforstretch, chae2008making, Lu2012opportunities}. For fibre spun from arrays of aligned CNTs (forests), tensile strength and modulus generally were observed to increase with increasing CNT length \cite{zhang2007ultrastrong}. This seems reasonable because longer tubes imply fewer tube-ends, which are regarded as defects \cite{zhu2011self}. Similarly, some reports have compared tensile properties of CNT fibres produced from different carbon precursors and thus composed of different constituent CNTs in terms of number of diameter and number of layers, spanning from single-walled (SWNT) to multi-walled (MWNT) CNTs \cite{motta2005mechanical,jia2011comparison}. Based on empirical evidence after fibre optimisation, there is consensus that large diameter few-layer CNTs result in superior fibre axial properties; a consequence of improved CNT packing and maximized contact area upon tube collapse \cite{Motta2017highperformance}. Extensive experimental work by Espinosa and co-workers has focused on multi-scale testing of CNT fibres and subunits to clarify the main factors limiting tensile strength \cite{beese2014key}. Thus, it was found the key role of CNT alignment on yarn performance, as well as the importance of interfacial strength and bundle strength in terms of different failure mechanisms.   

A fundamental difficulty to extract qualitative structure-property relations is to decouple the various interlinked features, such as CNT orientation, composition, length and association in crystals. Vilatela \textit{et al.} \cite{vilatela2011model} proposed a model for tensile strength of CNT fibres based on their yarn-like structure \cite{vilatela2010yarn} and fibrillar fracture. It considered an ensemble of parallel rigid rods, with load transfered by shear stresses between fibrous elements, the bundles, until a critical stress produced catastrophic failure by pull-out. Accordingly, fibre strength $\sigma$ was reduced to the product of total contact between load-bearing elements, their length $l$ and shear strength $\tau_F$.

\begin{equation}
\label{eq:equation11}
\sigma = \frac{1}{6}\;\Omega_1\; \Omega_2\; \tau_{F}\; l,
\end{equation}

\noindent where $\Omega_1$ is the fraction of the total number of graphene layers on the outside of the fibrous elements and $\Omega_2$ is the fraction of the outer graphene walls of the elements in contact with neighboring elements. This simple model captured the essense of the failure mechanism and its relation to fibre structure, providing fibre strength predictions (3.5 GPa/SG) in the range observed for small gauge length measurements (5 GPa/SG) \cite{koziol2007high}, but had several limitations, most notably the assumption of perfect CNT orientation. 

More recently, Wei et al introduced a modified shear-lag model that predicts fibre strength accounting for the length distribution of load-bearing elements, for both aligned and twisted CNT fibres \cite{wei2015new}.  Equipped with a Weibull distribution to take into account the probability of tensile fracture of CNT bundles, the Montecarlo-based model predicts upper bounds on CNT yarn mechanical properties very close to experimental values. The main limitations of this model are that it does not take into account the broad distribution of CNT orientation in real fibres; and its reliance on knowledge of bundle dimensions, which are difficult to determine accurately. However, simulations reveal different failure mechanisms in terms of bundle strength (which in turn is a function of the type of tubes and their arrangement), bundle length and interfacial strength.

In contrast with this plethora of incomplete descriptions, the fibrillar crystallite model developed originally for polymer fibres \cite{northolt1985elastic, northolt2005tensile} can successfully describe the mechanical properties of a wide range of materials, ranging from cellulose to high-performance fibres, including carbon fibres (CF) and rigid-rod polymer fibres. In this work, we show that macroscopic fibres of CNTs can also be treated as ensembles of fibrillary crystals, corresponding to bundles of individual tubes. By studying samples produced with controlled degree of alignment, we show that their tensile properties can be simply determined by the crystal shear strength and modulus and the orientation distribution of crystallites relative to the fibre axis. This provides accurate predictions of fibre modulus, strength and fracture envelope for a range of CNT fibres produced in-house with controlled alignement and compositions, as well as with others reported in the literature. \textit{In-situ} orientation measurements by synchrotron X-ray during tensile testing confirm the accommodation of axial deformation of CNT fibres by crystal stretching and rotation, the core idea of the model.

\section{Experimental}
CNT fibres were synthesized by the direct spinning method whereby an aerogel of CNTs is directly drawn out from the gas phase during growth by floating catalyst chemical vapor deposition (CVD) \cite{Li}. Two sets of fibres were produced, with differences in their constituent CNTs. fibres of few-layer MWNTs were synthesized using butanol as carbon source and adjusting the promotor (sulphur) content accordingly to produce CNTs with the desired number of layers \cite{reguero2014controlling}. fibres of collapsed DWNTs were produced using tolune as carbon precursor. In both cases ferroncene was used as iron catalyst source and Hydrogen as carrier gas. The reaction was carried out at 1250$^{\circ}$C in a vertical tubular furnace reactor. For each sample set, the degree of CNT orientation in the fibre was varied by changing the rate at which the fibres were drawn out of the reactor \cite{aleman2015strong}, equivalent to the winding rate.  

The mechanical properties of the different fibre samples were determined from tensile tests on individual CNT fibre filaments, using a gauge length of 20 mm and a strain rate of 2 mm/min. The tests were carried with a Textechno Favimat, equipped with a high-resolution 210 cN load cell. Fibre linear density was determined by weighing a know length of fibre (around 30m) and by using the vibroscopic method. 

Small- and wide-angle two-dimensional X-ray scattering patterns were obtained in the Non-crystalline diffraction (NCD) beamline at ALBA Synchrotron. The radiation wavelength was 1.0 {\AA} and the spot size at the focal plane of approximately 100 $\mu m$ $ X $ 50 $\mu m$. Sample-to-detector distance and other parameters were callibrated using reference materials. Data were processed with the software Dawn \cite{filik2017processing}. 

\textit{In-situ} tensile tests were performed on 15 mm samples using a Kammrath und Weiss miniaturized tensile stage. For these tests, strain steps were applied at a strain rate of 5 $\mu m$/s. SAXS patterns were acquired at fixed strain as the load was monitored.

Scanning Electron microscopy (SEM) was carried out with an FIB-FEGSEM Helios NanoLab 600i (FEI) at 10kV.TEM images were taken using a Talos F200X FEI operating at 20KV.
\section{The elastic extension of CNT fibres}

\subsection{The uniform stress model}
\label{sect:section21}

\begin{figure}[H]
\begin{center}
\begin{tabular}{c}
\includegraphics[width=\textwidth]{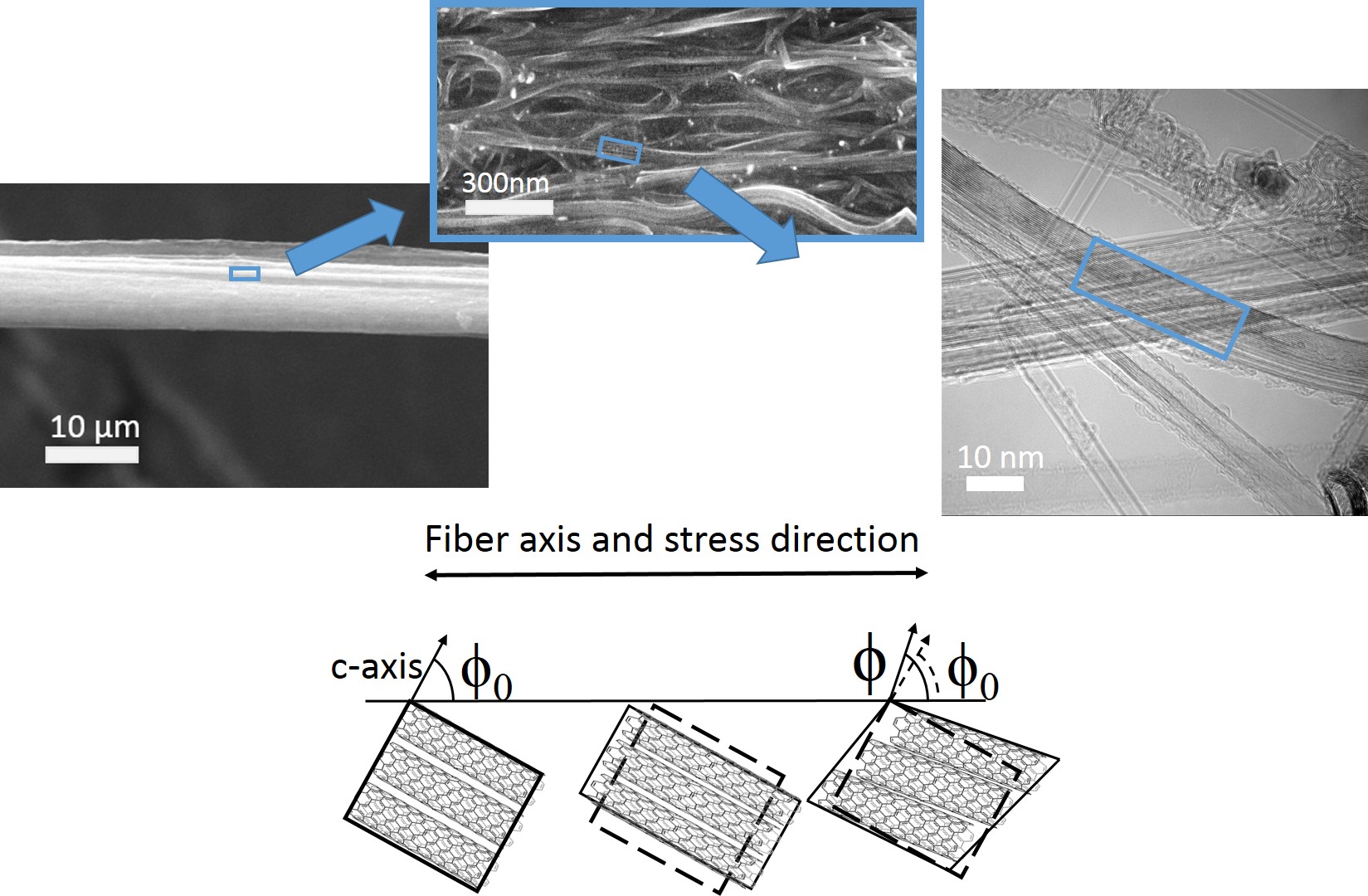} 
\end{tabular}
\end{center}
\caption 
{ \label{fig:Figure21}
(above) SEM and TEM images of a CNT fibre which reveal a fibrillar microestructure made up of close packed bundles of nanotubes $\&$ (below) schematic representation of a single CNT bundle and its contribution to the macroscopic deformation of the fibre: axial stretching by tensile deformation of nanotubes and crystallite rotation by shear between nanotube layers.} 
\end{figure}

The mechanical behavior of fibrous materials depends critically on their morphology \cite{Morton2008physical}. In this regard, despite the complex hierarchical structure found in CNT fibres \cite{yue2017fractal}, their microstructure can be defined as fibrillar by noting that CNT bundles are essentially long fibrils well-aligned along the fibre axis, as shown in the electron micrograph in Fig.\ref{fig:Figure21} (a). It is precisely these fibrils which act as load-carrying elements and, therefore, their mechanical properties control to a large extent the final properties of the macroscopic fibre.

Within the framework of a uniform stress model, a fibre is considered to be made up of an array of identical fibrils, which are all subjected to a uniform stress along the fibre axis \cite{ward1962optical,ward1983mechanical}. Following the analogy to polymers, it is also assumed that each fibril consists of crystallites arranged end-to-end. In the case of CNT fibres, the crystallites are bundles in which the CNTs are closed-packed and parallel to each other at a separation between that in Bernal and turbostratic graphite. These fibrils (bundles) are the basic structural elements in the fibres and therefore the key elements in our continuum mechanics analysis.

The structure outlined above is conceptually similar to that of carbon fibres, which have indeed been treated as networks of fibrillar crystalline domains formed by stacks of graphitic planes and defined by a symmetry axis, corresponding to the normal to the graphite basal plane (c-axis) \cite{northolt1991tensile}. Evidently, the orientation of crystallites in CNT fibres is also defined by the normal to the graphitic planes, that is, to the CNT main axis. For such well oriented fibres, it can be demonstrated \cite{northolt1991tensile} that their Young's modulus ($E$)  is given by 

\begin{equation}
\label{equation21}
\frac{1}{E} = \frac{1}{e_c}+\frac{<cos^2\phi_{0}>}{g}
\end{equation}

\noindent where $e_c$ is the modulus in the direction normal to the c-axis, $g$ is the shear modulus between planes oriented normal to the c-axis and the parameter $<cos^2\phi_{0}>$ is the second moment of the c-axis orientation distribution in the unloaded state, defined latter. According to this expression, there are two contributions from the crystallites to the fibre strain, as schematically depicted in Figure \ref{fig:Figure21}(b).  The first term refers to the axial elastic stretching of crystallites, that is to nanotubes themselves, whereas the second term involves the effect of crystallite alignment due to shear strain. This angular deformation (shear component) implies the rotation of nanotubes toward the fibre axis, increasing the angle between the c-axis and the fibre axis from $\phi_{0}$ to $\phi$. From this description, it is evident that the parameters $e_c$ and $g$ correspond to the Young's modulus of CNT bundles and the shear modulus associated to tangential elastic displacement between nanotubes, respectively.

\subsection{Orientation analysis by Small Angle X-Ray Scattering}
\label{sect:section22}

\begin{figure}[H]
\begin{center}
\begin{tabular}{c}
\includegraphics[width=0.9\textwidth]{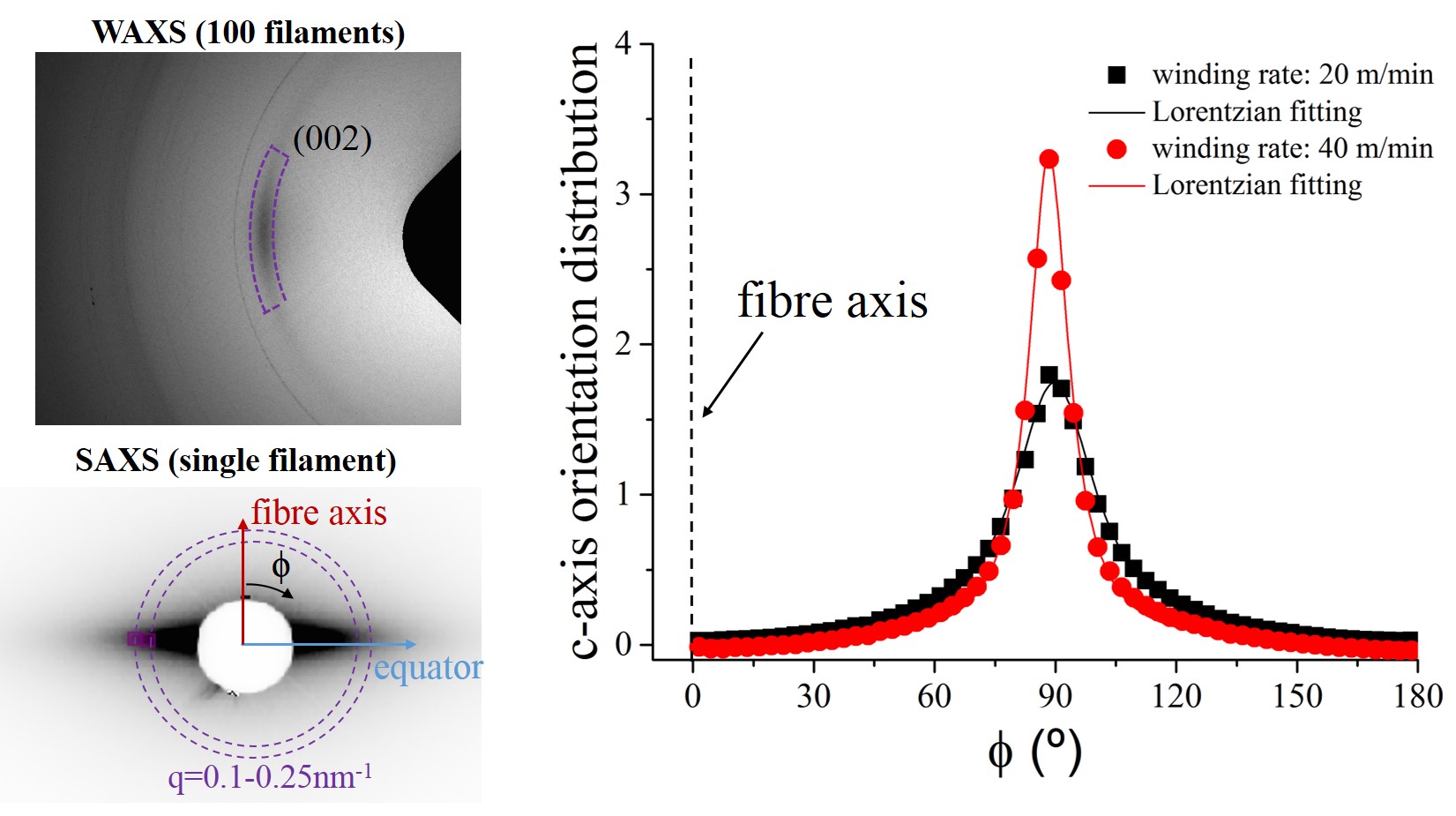}  
\\
\hspace{-2cm} (a) \hspace{5cm} (b)
\end{tabular}
\end{center}
\caption 
{ \label{fig:Figure23}
(a) WAXS (above) and SAXS (below) patterns of CNT fibres $\&$ (b) c-axis orientation distribution functions for two CNT yarns obtained at different draw ratios.} 
\end{figure}

As mentioned above, the structure of the fibre determines its modulus via the orientation distribution of crystallites, embedded in the term $<cos^2\phi_{0}>$ in equation \ref{equation21}. The orientation distribution function (ODF) of crystallites in fibres is typically obtained by 2D wide-angle X-ray scattering (WAXS) \cite{roe2000methods, martinezhergueta2015}.In the case of CNT arrays, this is done by analysis of the (002) interplanar reflection arising from adjacent CNTs as well as from internal CNT layers\cite{li2007x}, and giving rise to an equatorial feature in WAXS data as the one shown in Figure\ref{fig:Figure23}(a) \cite{severino2016progression,zhang2010nanocomposites,behabtu2013strong}. The ODF can be obtained from the azimuthal profile of scattering intensity obtained after radial integration, $I(\phi)$

\begin{equation}
\label{eq:equation22}
\Psi(\phi) = \frac{I(\phi)}{\int_{0}^{\pi} I(\phi) sin(\phi) d\phi}, 
\end{equation}

Note that this orientation distribution function is of the normal to the graphene basal planes in the crystallites and thus perpendicular to the CNT main axis. 

With knowledge of the ODF $<cos^2\phi>$ can be calculated by averaging $cos^2\phi$ over the c-axis orientation distribution as

\begin{equation}
\label{eq:equation23}
<cos^2\phi> = \int_{0}^{\pi}  cos^2\phi \; \Psi(\phi) \; sin\phi \; d\phi
\end{equation}

Because of the weak X-ray scattering of CNT fibres, WAXS measurements are typically carried out on multiple filament samples and using synchrotron X-ray sources. However, we have recently shown that such samples have an intrinsically high misorientation between filaments relative to the intrinsic fibre orientation. These makes them unsuitable to determine the ODF of individual fibres. Instead, it is more accurate to use SAXS, which because of its higher intensity can be readily measured on individual fibres in standard synchrotron radiation facilities. Figure\ref{fig:Figure23}(a) shows an example of a 2D SAXS pattern from an individual CNT fibre. The equatorial streak is characteristic of fibrillar structures in high-performance fibres such as PBO \cite{ran2002situ}, Kevlar \cite{dobb1979microvoids} or carbon fibre \cite{gupta1994small}. The use of SAXS data instead of WAXS is possible because unlike other CNT arrays\cite{meshot2017quantifying}, for CNT fibres the orientations measured from WAXS and SAXS are equivalent over a wide scattering vector ($q$) range \cite{davies2009structural}(Suplementary Information). In CNT fibres SAXS arises mainly from the network of elongated mesopores and bundles in the fibre, which corresponds precisely to the orientation of interest for this work. A further point of interest is that both the WAXS and SAXS azimuthal profiles are best fit by a Lorentzian, rather than a Gaussian distribution. The Lorentzian profile intrinsically leads to low values of Herman's parameter (0.5) even for highly oriented fibres, and cannot therefore be taken as a direct indicator of high-performance fibre properties. 

With the aim of understanding orientational effects on tensile properties, in this study we have produced fibres with different degree of CNT orientation, obtained by varying their drawing rate during fabrication  \cite{aleman2015strong}. This effect is clearly seen in the SAXS ODF plotted in Figure \ref{fig:Figure23}(b). Clearly, the sample produced at a higher draw ratio has a narrower ODF. In addition, we have also prepared samples synthesised from two different precursors (see experimental details), and which have differences in their constituent CNTs and tensile properites. Once consists predominantly of collapsed double-walled carbon nanotubes (DWNT) and the other of few-layer (3-5) multi-walled carbon nanotubes (MWNT). The properties of these samples are summarised in Table \ref{table:table11}.

\begin{table}[H]
\centering
\caption{Experimental values of CNT fibres}
\label{table:table11}
\begin{tabular}{c|c|c|c|c|c|}
\cline{2-6}
                                       
                                      &\begin{tabular}[c]{@{}c@{}}winding rate \\(m/min)\end{tabular} &\begin{tabular}[c]{@{}c@{}}$<cos^2\phi_{0}>$ \\($\times 10^\textsuperscript{-2}$)\end{tabular}  & \begin{tabular}[c]{@{}c@{}}$E$ \\(GPa)\end{tabular} & \begin{tabular}[c]{@{}c@{}}$\sigma_b$ \\(GPa)\end{tabular}  & \begin{tabular}[c]{@{}c@{}}Fracture energy \\(J/g)\end{tabular} \\ \hline
\multicolumn{1}{|l|}{\multirow{4}{*}{\begin{tabular}[c]{@{}c@{}}Collapsed \\DWNTs\end{tabular}}} &4 &8.89  & $44 \pm 9$ & $1.0 \pm 0.2$  & $70 \pm 40$ \\ \cline{2-6} 
\multicolumn{1}{|l|}{}                  &8 & 9.7  &$32 \pm 7$  &$1.1 \pm 0.1$  & $90 \pm 20$  \\ \cline{2-6} 
\multicolumn{1}{|l|}{}                  &12 & 7.46 & $56 \pm 8$ & $1.3 \pm 0.2$ & $70 \pm 30$  \\ \cline{2-6} 
\multicolumn{1}{|l|}{}                  &16 & 5.42 & $61 \pm 7$ &$1.7 \pm 0.3$  &$100 \pm 30$  \\ \hline
\multicolumn{1}{|l|}{\multirow{3}{*}{\begin{tabular}[c]{@{}c@{}}Few-layer \\MWNTs\end{tabular}}} &20 & 11.58 & $33 \pm 8$ & $0.7 \pm 0.1$   & $60 \pm 10$ \\ \cline{2-6} 
\multicolumn{1}{|l|}{}                  &30 & 10.08  & $38 \pm 8$ & $0.8 \pm 0.1$ & $65 \pm 15$ \\ \cline{2-6} 
\multicolumn{1}{|l|}{}                  &40 & 6.37 & $64 \pm 16$ &$1.1 \pm 0.2$  & $80 \pm 40$  \\ \hline
\end{tabular}
\end{table}

\subsection{Results and discussion}
\label{sect:section23}

\begin{figure}[H]
\raggedright
\begin{tabular}{c}
\includegraphics[width=0.9\linewidth]{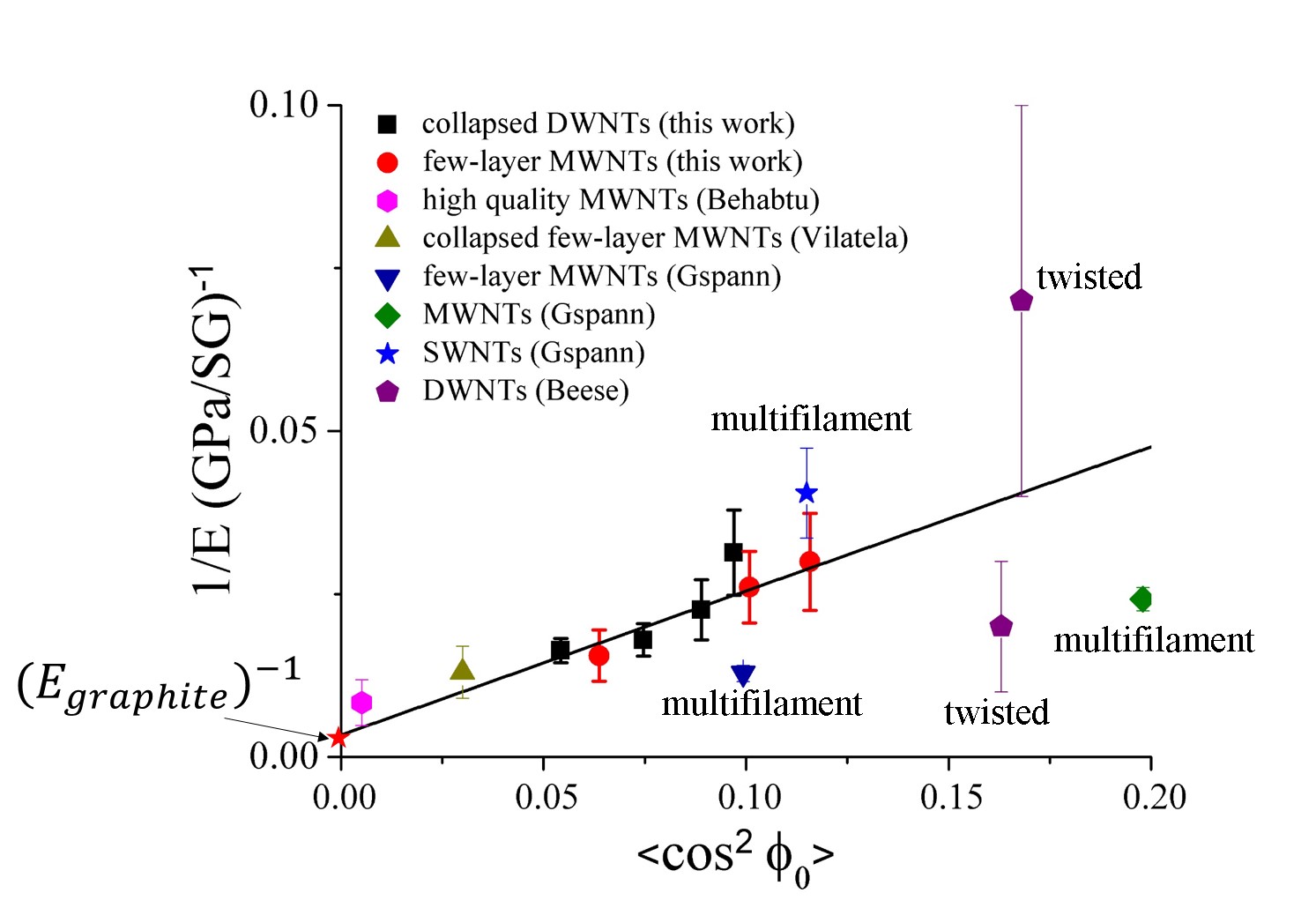}
\end{tabular}
\caption 
{ \label{fig:figure24}
Compliance  ($E\textsuperscript{-1}$) of CNT fibres plotted against the orientation parameter $<cos^2\phi_{0}>$ for both in-house fibres and data obtained from the literature. These last correspond to both single \cite{vilatela2011structure,behabtu2013strong} and multifilament samples\cite{gspann2017high}. In addition, data from twisted CNT yarns have been also plotted \cite{beese2014key}. The linear fit is for CNT fibres produced and analysed in this work.}
\end{figure}

Figure \ref{fig:figure24} presents values of fibre compliance ($E\textsuperscript{-1}$) plotted against the orientation parameter $<cos^2\phi_{0}>$ determined from SAXS. As can be observed, the experimental data exhibit a linear correlation with the orientation parameter, which is in excellent agreement with equation \ref{equation21} and support the use of the uniform stress transfer model for oriented fibres. The linear fit includes data for CNT fibres with different CNT types and different tensile properties, as discussed before. Moreover, literature data also follow the same trend (see supplementary material for a discussion of literature data). This behaviour is extremely relevant, because it naturally leads to the conclusion that the stiffness of a CNT fibre is mainly dominated by crystallite alignment, represented here by the parameter $<cos^2\phi_{0}>$. It also implies that for fibres with constituent CNTs with few layers (1-5), the internal layers of the CNTs make a substantial contribution to the fibre stiffness. 

Assuming a CNT fibre specific gravity of 1.8, values for $e_c\approx 540$ GPa and $g= 8.1 \pm 1.8$ GPa are obtained from the fitting. The crystallite stiffness value $e_c$ is close to the in-plane Young's modulus of graphite ($E=1020 \pm 30$ GPa)\cite{morgan2005carbon} and in the range of experimental values for individual CNTs and bundles \cite{peng2008measurements,yu2000tensile}. Considering the spread in fibre stiffness values, the agreement is remarkable. The value for $g$ is above the theoretical shear modulus of single-crystal graphite (G\textsubscript{graphite}=4.6 GPa)\cite{nicklow1972lattice}. However, measurements on carbon fibres (CF) with a more complex polycrystalline structure \cite{johnson1987structure}, \cite{northolt1991tensile} give values of 5 to 33 GPa on account of out-of-plane interactions arising from crystallite edges and defects, typical of graphitic ensembles with a wide distribution of interlayers spacings. Morphology and contact area in CNT fibres are different from Bernal graphite and thus a compact arrangement of parallel CNTs needs not display the same response in terms of stress than a standard stack of graphitic planes when subjected to shear strain.  Therefore, the value reported here, $g= 8.1 \pm 1.8$ GPa, can be considered a reasonable, conservative estimate for the shear modulus of CNT crystallites.

\section{Fracture model: CNT fibre as a molecular composite}
\label{sect:fracture} 

\begin{figure}[H]
\begin{tabular}{c}
\includegraphics[width=\linewidth]{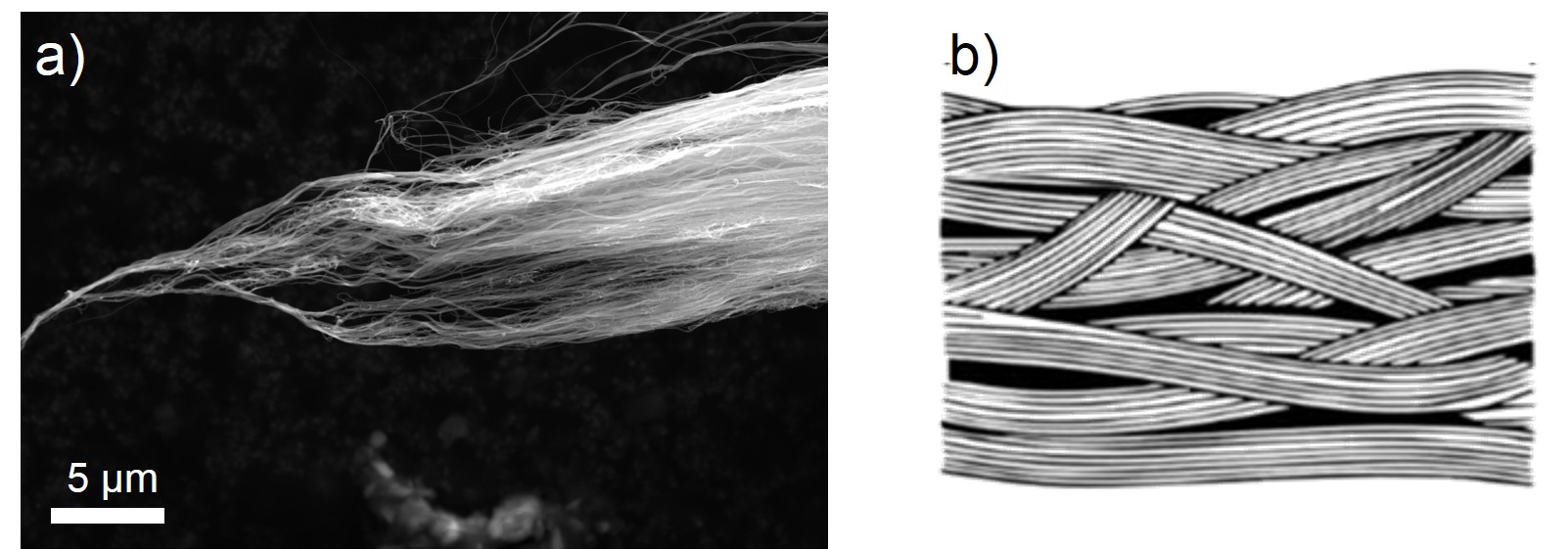}
\end{tabular}
\caption 
{ \label{fig:figure25}
a) SEM image of the fracture surface of a CNT fibre in which a fibrillar morphology is revealed $\&$ b) schematic structure of the CNT fibre as a network of well-aligned CNT bundles.}
\end{figure}

The model discussed so far successfully describes only the elastic axial deformation of CNT fibres. In order to  provide a description of factor governing tensile strength we first considering a CNT fibre as a composite of strong/stiff crystallites in a matrix of weak secondary bonds. This approach describes the tensile strength of  fibres such as aramid, which have a highly fibrillar fracture analogous to that of a uniaxially oriented fibre-reinforced composites that fail in tension via matrix shear failure initiated at the fibre ends \cite{knoff1987relationship}. The fracture mechanism in CNT fibres is indeed fibrillar \cite{vilatela2011model}. As shown in Figure \ref{fig:figure25}, the fracture ends of the fibre shows failure by extensive shear-induced decohesion between CNT bundles.

When treating a high-performance fibre as a molecular composite of filaments in a matrix of secondary bonds, fibre strength can be obtained from a modified form of the Tsai-Hill criterion for failure in uniaxial composites \cite{northolt2005tensile}. According to it, the strength of a composite loaded in a direction at an angle $\theta$ with respect to the parallel aligned fibres is given by 

\begin{equation}
\label{eq:equation31}
\sigma\textsubscript{comp}=[\frac{cos^4\theta}{\sigma_L^2}+(\frac{1}{\tau_b^2}-\frac{1}{\sigma_L^2})\; sin^2\theta \; cos^2\theta + \frac{sin^4\theta}{\sigma_T^2}]^{-\frac{1}{2}}
\end{equation}

\noindent where $\sigma_L$ is the axial strength of the fibres, $\sigma_T$ is the strength normal to the composite's symmetry axis, and $\tau_b$ is the critical shear strength in a plane parallel to the fibres \cite{hull1996introduction}.

Such a model can easily be applied to a CNT fibre by introducing the average angle between the axial loading direction and load-bearing elements $<cos^2\phi>$, obtained from the ODF at fracture ($\Psi(\phi_{b})$) and noting that $\phi=\frac{\pi}{2}-\theta$. In addition, for highly aligned fibres the transverse properties are negligible and the last term in \ref{eq:equation31} can be neglected. Finally, fibre tensile strength can be approximated by the expression 

\begin{equation}
\label{eq:equation32}
\sigma_b\approx[\frac{<sin^4\phi_b>}{\sigma_L^2}+(\frac{1}{\tau_b^2}-\frac{1}{\sigma_L^2})\; <sin^2\phi_b \; cos^2\phi_b>]^{-\frac{1}{2}}
\end{equation}

This expression contains two unkown parameters: the critical axial fibril strength $\sigma_L$ and the critical shear strength $\tau_b$. In a macrocomposite the fibres are continuous and $\sigma_L$ is simply the strength of fibres. But in the case of a CNT fibre visualised as a composite, its constituent fibrils (the crystallites) are of finite length, which implies that load is transfered from one fibril to another through shear (shear lag). Shear stress arising at the end of a filament can, upon exceeding a limiting stress $\tau_b$, cause debonding of the filament from its nearest neighbors. $\sigma_L$ is thus the \textit{maximum axial stress} in the fibrils before shear failure. 
$\sigma_L$ is clearly then dependent on the shear strength $\tau_b$. In this regard, Yoon\cite{yoon1990strength} derived an expression for a polymer fibres of very long chains and failure in shear, which relates these two parameters with the crystallite elastic constants $e_c$ and $g$. Applied to CNT fibres it leads to 

\begin{equation}
\label{eq:equation36}
\sigma_L=1.14\cdot\tau_b\cdot\sqrt{\frac{e_c}{g}},
\end{equation}

Implicit in the model discussed above is the view that the CNT fibres is treated as a network of fibrils, corresponding to long crystalline domains, that is crystallites, similar in cross section to a bundle. Failure occurs through shearing of crystallites, leading to fibrillar fracture before any CNT rupture occurs. The network is a continum of crystalline domains and there is no reason to expect that all ends of CNTs in a domain match and hence that a bundle terminates abrutly. Instead, the 1-mm long CNTs can easily form part of several crystalline domains, very much in the same way polymer chains do.

\subsection{Relation between Young's Modulus and ultimate strength}
\label{sect:section32}

The model can be contrasted with experimental data by relating the fibre modulus and strength, with the use of equations \ref{eq:equation32} and \ref{equation21}. In the process, it is necessary to determine the ODF at the point of fracture $<cos^2\phi_{b}>$. In the uniform stress model, upon fibre loading, crystallites deform \textit{elastically} in shear and re-align parallel to the fibre axis. The second moment of c-axis orientation distribution $<cos^2\phi>$ of a fibre under a stress $\sigma$ decreases with respect to the unloaded state $<cos^2 \phi_{0}>$ according to the following expression \cite{northolt1991tensile}:

\begin{equation}
\label{eq:equation37}
<cos^2\phi>=<cos^2\phi_{0}>\exp{(-\frac{\sigma}{g})}
\end{equation}

\noindent where $g$ is the crystallite shear modulus as described before. 
In the case where fracture involves additional crystallite alignment through plastic deformation by shear, this expression can be modified to obtain \cite{northolt2005tensile}  

\begin{equation}
\label{eq:equation38}
<cos^2\phi_b>=<cos^2\phi_{0}>\exp{(-\frac{\sigma_b}{g_v})}
\end{equation}

\noindent where $<cos^2\phi_b>$ corresponds to the orientation at fracture, as in equation \ref{eq:equation32}.

\begin{figure}[H]
\begin{center}
\begin{tabular}{c}
\includegraphics[width=\textwidth]{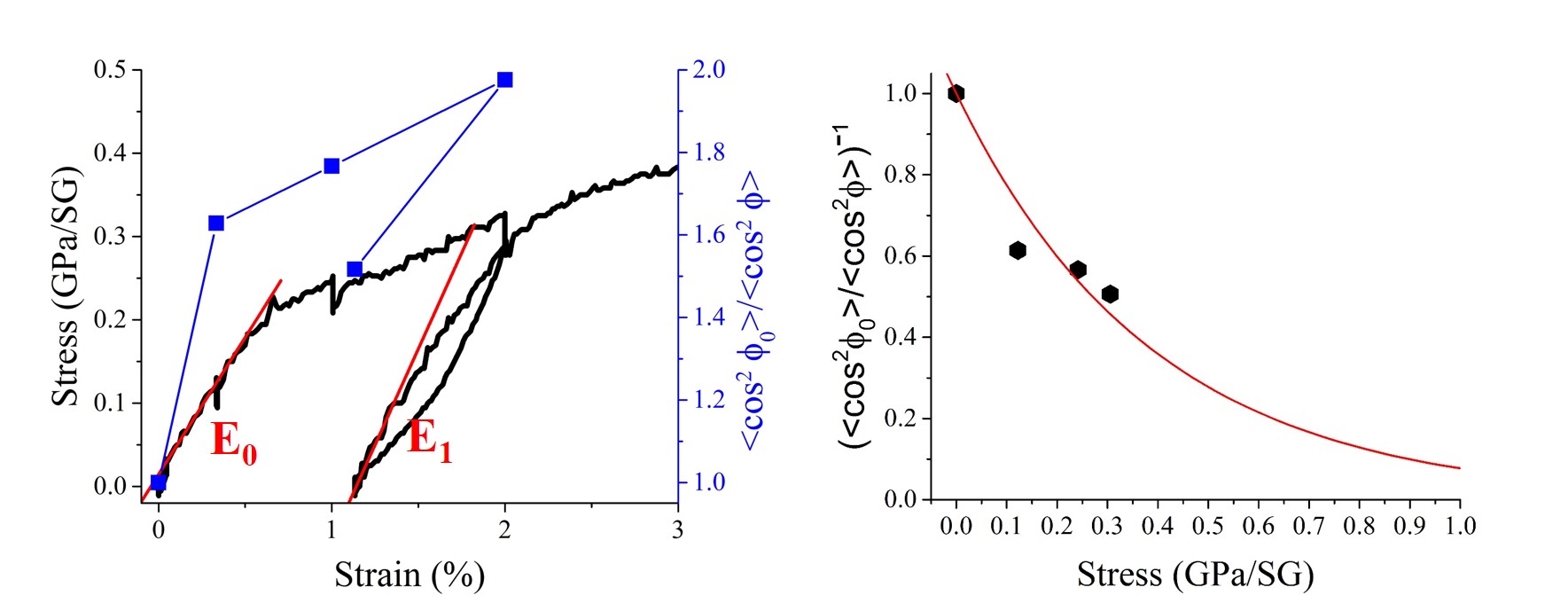}
\end{tabular}
\end{center}
\caption 
{ \label{fig:figure33}
(left)  Results obtained from subjecting a CNT fibre to an \textit{in situ} tensile test with SAXS measurements. Thus, both the stress and the parameter   $\frac{<cos^2\phi_{0}>}{<cos^2\phi>}$, calculated from SAXS measurements, are plotted against strain. (Right) Evolution of the parameter  $\frac{<cos^2\phi_{0}>}{<cos^2\phi>}$ with the stress during tensile stretching and their accurate fitting to the expresion. }
\end{figure}

We have measured the evolution of the ODF by in-situ SAXS measurements during tensile deformation of a CNT fibres. The data, shown in Figure \ref{fig:figure33}, confirm the exponential relation in equation \ref{eq:equation38}. It gives a value for $g_v = 0.7 GPa$, discussed further below. The stress-strain curve confirms that after plastic deformation upon reloading the fibre has a higher modulus due to a higher degree of orientation, as observed for example in rigid-rod high-performance polymer fibres \cite{northolt1985elastic}, expressed in quantitative terms as $\frac{E_1}{E_0}\approx\frac{<cos^2\phi_{0}>}{<cos^2\phi>}$

Equipped with the relation between the ODFs and equations \ref{equation21}, \ref{eq:equation32}, and \ref{eq:equation38}, and using trigonometric approximations assuming that the misalignment is small, we obtain an expression relating the Young's modulus with the ultimate tensile strength (Supplementary Information):

\begin{equation}
\label{eq:equation39}
\frac{1}{E} = \frac{1}{e_c}+\frac{exp(\frac{\sigma_b}{g_v})}{g} \cdot \frac{(3\; \sigma_L^{-2}-\tau_b^{-2})+\sqrt{(\tau_b^{-2}-3\; \sigma_L^{-2})^2-4\; (4\; \sigma_L^{-2} - 2\; \tau_b^{-2})\; (\sigma_L^{-2}- \sigma_b^{-2})}}{2\; (4\; \sigma_L^{-2}- 2\; \tau_b^{-2})}
\end{equation}

\begin{figure}[H]
\begin{center}
\begin{tabular}{c}
\includegraphics[width=0.9\textwidth]{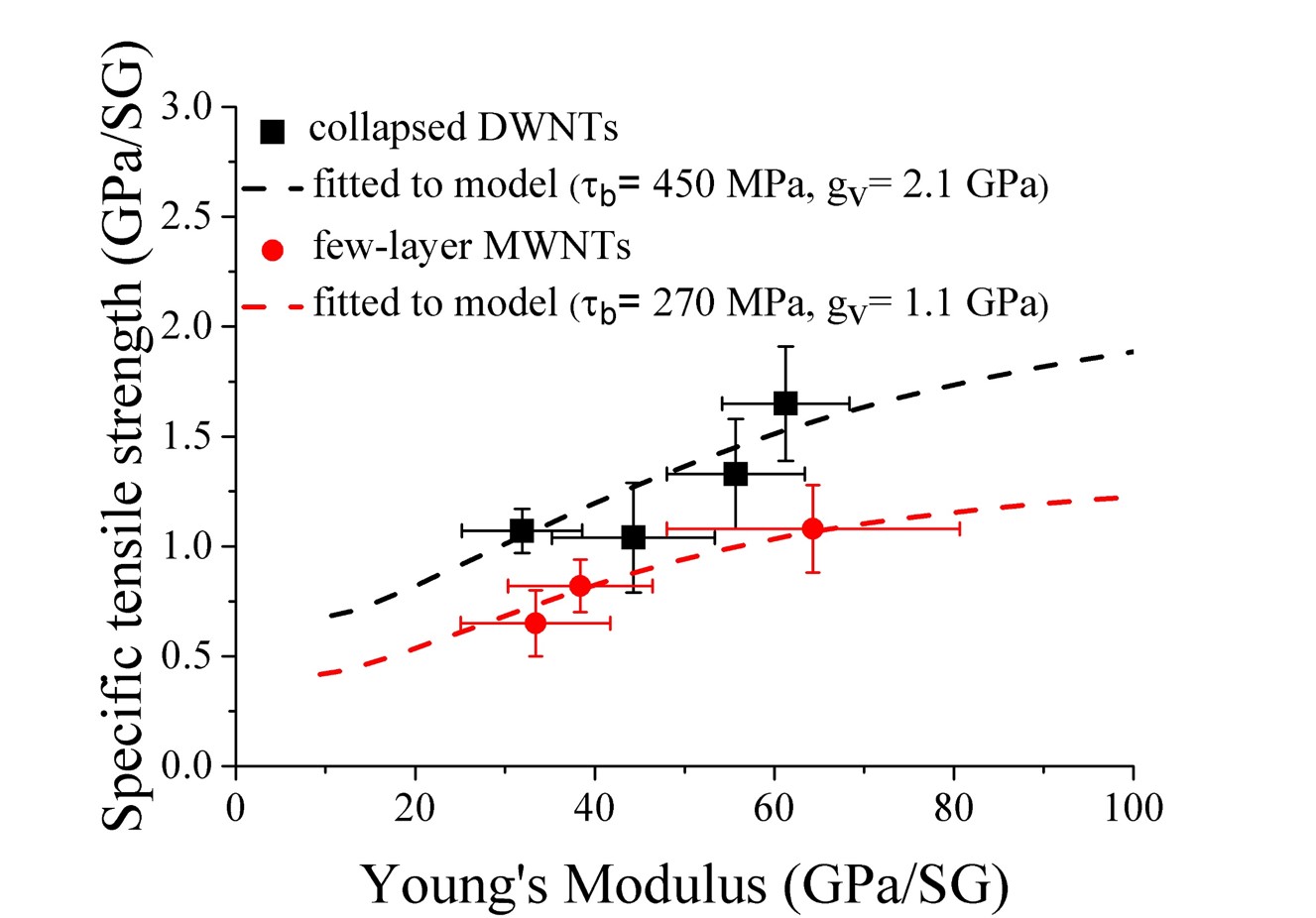}
\end{tabular}
\end{center}
\caption 
{ \label{fig:figure32}
Graph of tensile strength against tensile stiffness for each butanol and toluene made CNT fibres. The modified Tsai-Hill failure criterion for polymer fibres fits accurately in both cases (dashed-lines). This model enables to fit parameters $\tau_b$ and $g_v$ which define maximum shear strength and shear stiffness of the crystallites, respectively.}
\end{figure}

In equation \ref{eq:equation39}, only $\tau_b$ and, in principle $g_v$, are unknowns, but both can be obtained by fitting experimental data. As shown in Figure \ref{fig:figure32}, there is a very good match between experimental data and the fitting to equation \ref{eq:equation39} for both types of CNT fibres. The extracted value of $g_v = 1.1$ GPa for few-walled MWNT fibres is close to the that determined by in-situ SAXS ($\approx 0.7$ GPa). Further success of this expression is the prediction that both fibre strength and modulus increase with improved alignement, as observed experimentally a decade ago \cite{koziol2007high}. Additionally, with this plot at hand the differences in properties of the two sets of fibres can be now ascribed to different values of $\tau_b$ and $g_v$.

\section{The fracture envelope}
\label{sect:section4}

The model accuracy is tested again by determining the fracture envelope of the CNT fibres; that is, the set of stress-strain coordinates where the fibre fails. For brittle linear-elastic fibres such as CF, the stress-strain relation is \cite{northolt1991tensile}

\begin{equation}
\label{eq:equation391}
\epsilon \approx\; \frac{\sigma}{e_c}+<cos^2\phi_{0}>[1-exp(-\frac{\sigma}{g})]
\end{equation}

\begin{figure}[H]
\begin{center}
\begin{tabular}{c}
\includegraphics[width=\textwidth]{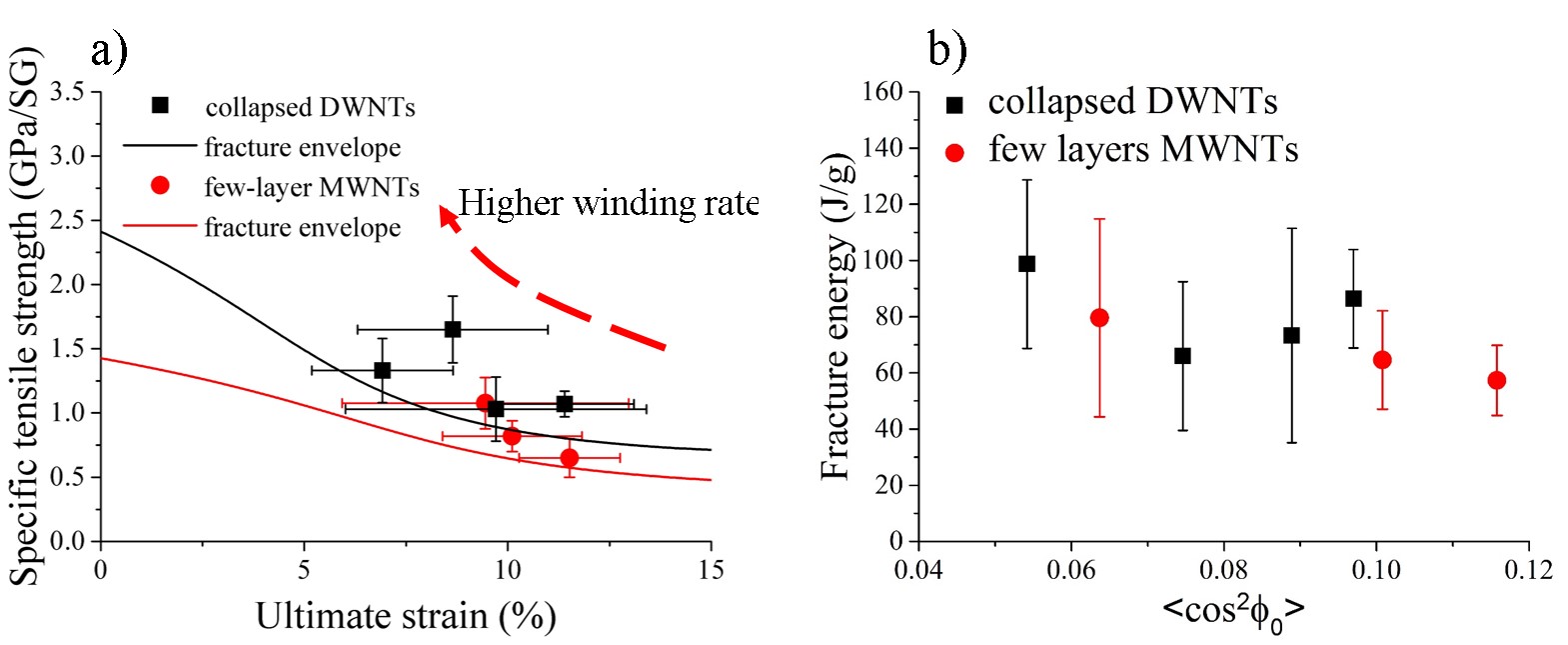}  
\\
\end{tabular}
\end{center}
\caption 
{ \label{fig:Figure41}
(a) Calculated fracture envelope compared to CNT fibres tested in this work $\&$ (b) fracture energy exhibited for each CNT fibre plotted against the initial orientation parameter $<cos^2\phi_{0}>$ } 
\end{figure}

\noindent In order to account for some plastic deformation in CNT fibres we replace  $g$ with $g_v$. Using then equation \ref{equation21}, we obtain the following expression for the strain-to-break:

\begin{equation}
\label{eq:equation392}
\epsilon_b \approx\; \frac{\sigma_b}{e_c}+g\; (\frac{1}{E}-\frac{1}{e_c})\; [1-exp(-\frac{\sigma_b}{g_v})]
\end{equation}

\noindent with the parameters in equation\ref{eq:equation392} obtained as discussed  before. This leads to the fracture envelope of stress-strain failure pairs ($\sigma_b$,$\epsilon_b$) for different fibre orientations. $Figure\;\ref{fig:Figure41}$ shows good agreement between experimental data and the predicted envelope for both types of CNT fibres. 

However, the fracture envelope seems to lie below the experimental values, particularly for fibres subjected to higher draw ratios, and therefore more oriented. In addition, the experimental data show a steeper decrease of $\sigma_b$ with strain than the prediction, an effect that cannot merely be explained in terms of the accuracy of our values for $g$ and $g_v$. Instead, it is likely that $g_v$ is not constant, but has a small dependance on the degree of alignment. This is the case for some polymer fibres, whose deformation mechanisms substantially depend on draw ratio \cite{northolt2005tensile}. In this respect, we note that CNT fibres subjected to higher draw ratios have a greater fracture energy ((Figure \ref{fig:Figure41}b), whereas the fibrillar breakage model assumes this to be constant through a constant number of failing elemets. Our recent WAXS measurements on multifilament samples suggest that samples produced at higher draw ratios have a larger "degree of crystallinity", that is, a large fraction of graphitic planes at turbostratic separation in coherent domains \cite{yue2017fractal}, which might be responsible for the this increase in fracture energy and the small deviation from the predicted fracture enevelope.  

\subsection{Comparison of fibres}
\label{sect:section5}

\begin{table}[H]

\begin{center}
\caption{parameters $\tau_b$ and $g_v$ determined for the model. }
\label{tab:table2}
	
    \renewcommand{\thefootnote}{\thempfootnote} 
    \begin{tabular}{| c | c | c | c | c |}
    \hline
     & \begin{tabular}[c]{@{}c@{}}$\tau_b$\\(MPa)\end{tabular} & \begin{tabular}[c]{@{}c@{}}$g_v$\\(GPa)\end{tabular} & \begin{tabular}[c]{@{}c@{}}$g_v/g$\end{tabular} & \begin{tabular}[c]{@{}c@{}}$\sigma_L$\\(GPa)\end{tabular}\\ \hline
    collapsed DWNTs & 450 & 2.1 & 0.26 & 5\\ \hline
    few-wall MWNTs & 270 & 1.1 & 0.12 & 2.5\\ \hline
    PpTA (Kevlar) & 370 & 1.2 & 0.7 & 4.87 \\ \hline
    PBO (Zylon) & 400 & 2.0 & 1.0 & 7.7 \\ \hline
    POK & 300 & 0.5 & 0.3 & 4.83 \\ \hline
    Cellulose II & 325 & 1.5 & 0.6 & 2.25 \\ \hline
    PET & 290 & 0.7 & 0.5 & 3.12\\ \hline
    HM50 (carbon fibre) \textsuperscript{\emph{a}} & 310 & 10 & 1.0 & 2.5\\ \hline
    \multicolumn{1}{}{}
    \textsuperscript{\emph{a}} estimated from Northolt \textit{et al.}\cite{northolt1991tensile}

    \end{tabular}
   
\end{center}

\end{table}

\begin{figure}[H]
\begin{center}
\begin{tabular}{c}
\includegraphics[width=\textwidth]{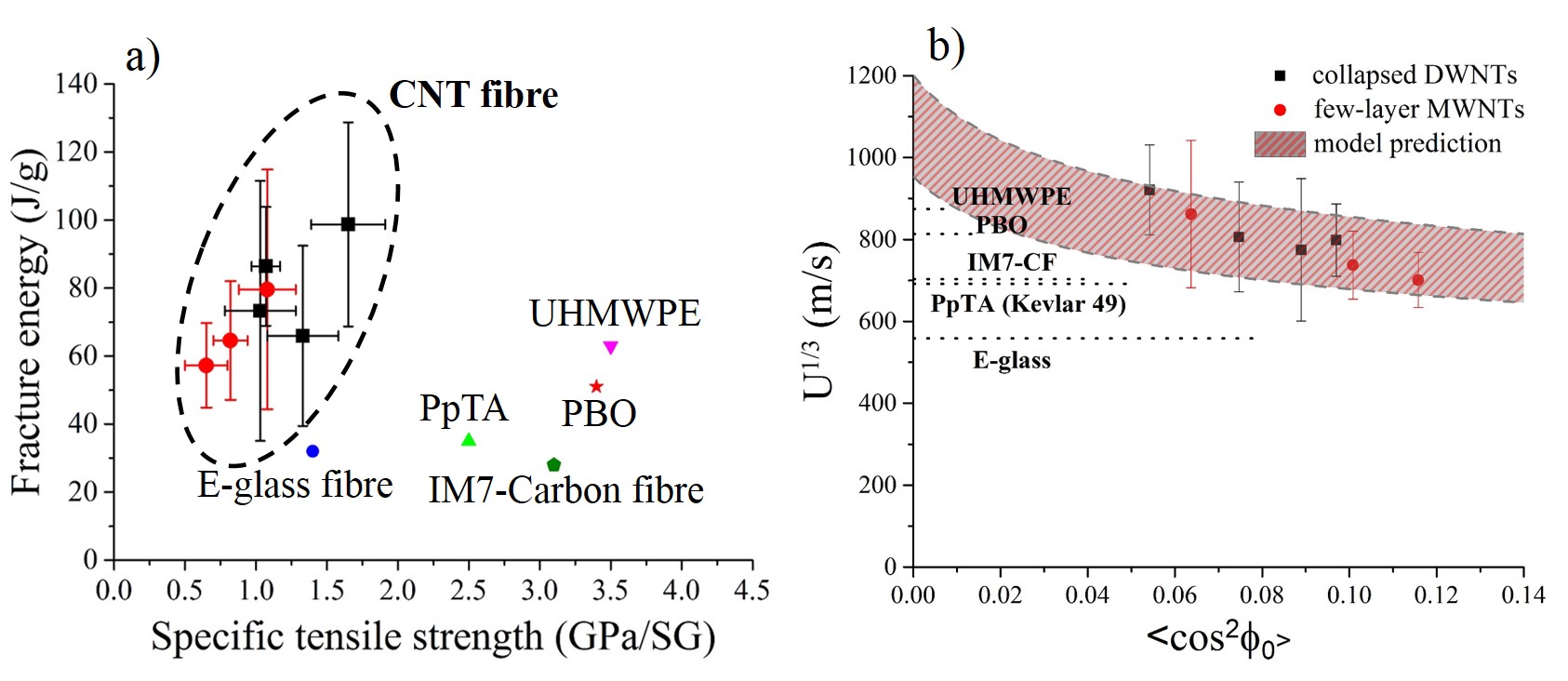}
\end{tabular}
\end{center}
\caption 
{ \label{fig:figure51}
(a) Fracture energy versus strength of high-performance fibers and of samples in this work and (b) prediction of the ballistic figure of merit $U^\frac{1}{3}$ of CNT fibers as a function of CNT orientation ($<cos^2\phi_{0}>$) showing superior properties than conventional high-performance fibers.}
\end{figure}

In Table \ref{tab:table2} we compare the parameters extracted from the fibrillar crystallite analysis for different fibres, including our two types of CNT fibres, CF, high-performance polymer fibres and ductile polymer fibres \cite{northolt2005tensile,northolt1991tensile}.  The values of $\tau_b$ and $g_v$ are in the same range for all the fibres, irrespective of their chemistry and the nature of the interaction between molecular building blocks. This suggests that $\tau_b$ and $g_v$ take the form of effective shear strength and modulus of the crystallite ensemble, and which can therefore not be easily reduced to the properties of a single-crystal. Nevertheless, the high values of $\tau_b$ contrast with the lubricity of graphite and the reported shear strength of measured on individual CNTs spanning from 0.04 MPa \cite{kis2006interlayer} to 69 MPa \cite{suekane2008static}. In this regard, we note that the length of the CNTs in this work is aproximately 1mm, which contrasts with the small length used in individual tests, and which could imply a larger contribution from defects \cite{paci2014shear}, surface impurities \cite{naraghi2010multiscale}, domains in crystallographic registry \cite{Jamessliding} and mechanical entanglements. Nevertheless, in spite of the relatively high values of $\tau_b$ the maximum axial stress in the fibrils, $\sigma_L$, is still much lower than the tensile strength of individual CNTs, leaving ample room for interfacial chemistry strategies \cite{EndoXlink} to improve shear stress transfer and thus increase fibre tensile strength.     
The parameter $g_v$ is clearly a critical one. It describes the shear deformation stiffness of crystallites, including their rotation towards the fibre axis when bearing load. It is not a conventional elastic shear modulus, but rather a secant shear modulus, involving both elastic and plastic deformation. Although its relation to the fundamental properties of a graphite single-crystal is still unclear, it is a convenient parameters to describe the extent of CNT realignment upon axial fibre loading. For more brittle, essentially linear elastic CNT fibres, $g_v$ tends towards $g$. But for the ductile fibres tested in this work, the comparison in Table \ref{tab:table2} shows that CNT fibres have a very low ratio of $g_v/g$. We think that this parameter is responsible for the unusual combination of high fracture energy and tensile strength in these CNT fibres ($Figure\;\ref{fig:figure51}$a). Embodied in it is the ability of the CNT crystalline network to undergo substantial reorientation upon loading, which seems to be a unique feature of CNT fibres. 
Combined high specific strength/modulus and energy to break are particularly relevant for impact resistant structures. A figure of merit for fibre ballistic protection, for example, is the cubed root of the product of sonic modulus and specific energy to break (\textit{T}), $U^\frac{1}{3}={(\sqrt{\frac{E}{\rho}}\;T)}^\frac{1}{3}$ \cite{cunniff2002high}. The samples in this work have average values of $U^{1/3}$ of 800 $\frac{m}{s}$ which is superior to most high-performance fibers, including aramid and carbon fibers. But more importantly, as Figure \ref{fig:figure51}b shows, the model introduced here predicts that very modest improvements in CNT orientation would lead to a fibre with unrivaled properties for ballistic and impact protection $U^\frac{1}{3}> 1000 m/s$ (see supplementary material), outperforming the best synthetic fibers available (ultrahigh molecular weight polyethylene (UHMWPE) and poly(p-phenylene-2,6-benzobisoxazole) (PBO)).

\section{Conclusions}
This works presents an analytical model to describe the tensile properties of fibres of CNTs. It assumes that their structure can be treated as a network of oriented crystallites, similarly to a high-performance polymer fibre, defined by the crystallite orientation distribution function and shear modulus and shear strength. Experimental values of initial ODF and tensile modulus show remarkable agreement with the model for fibres produced in-house with different constituent CNTs and for different draw ratios, as well as with literature data. 
By considering the CNT fibre as composite of stiff fibrils (crystallites) in a matrix of secondary bonds, we introduce expressions for tensile strength based on fibre-reinforced composite lamina theory. Plastic deformation through CNT crystallite reorientation is introduced via a secant shear modulus. Its predicted value based on statistical fibre strength/modulus data matches an experimental value determined from in-situ synchrotron SAXS measurements of the ODF during tensile testing.
Overall, the model provides a solid framework for the study of CNT fibres produced under different conditions, capable of separating orientational from compositional effects. Amongst future improvements to the model we highlight: elucidating the effects of crystallite size and role of CNT layers, a more robust physical interpretation of $g_v$ and its dependance on fibre orientation, and the prediction of CNT fibre properties embedded in polymer matrices. Work towards these improvements is in progress. 
The model provide a quantitative prediction of the effect of improvements in CNT orientation and shear stress transfer on fibre tesile properties. It shows that small improvements in orientation, for example, would lead to higher specific strength than aramid and most carbon fibres, and superior ballistic protection that any other synthetic high-performance fibre.

\section*{References}

\bibliographystyle{elsarticle-num}

\bibliography{bibliography}

\end{document}